\documentstyle{aipproc}

\begin{document}
\title{Young Collapsed Supernova Remnants:\\ Similarities and Differences\\
 in Neutron Stars, Black Holes,\\ and More Exotic Objects}

\author{James S. Graber$^*$ }
\address{$^*$407 Seward Square SE\\ 
Washington, DC 20003}

\maketitle

\begin{abstract}
Type Ia supernovae are thought to explode completely, leaving no condensed remnant, only an expanding shell. 
 Other types of supernovae are thought to involve core collapse and are expected to leave a condensed remnant, 
which could be either a neutron star or a black hole, or just possibly, something more exotic, such as a quark or 
strange star, a naked singularity, a frozen star, a wormhole or a red hole.  It has proven surprisingly difficult to determine which 
type of condensed remnant has been formed in those cases where the diagnostic highly regular pulsar signature 
of a neutron star is absent. We consider possible observational differences between 
the two standard candidates, as well as the more speculative alternatives. 

We classify condensed remnants according to whether they do or do not possess three major features:
1)a hard surface, 2)an event horizon, and 3)a singularity.  Black holes and neutron stars differ on all three criteria. 
 Some of the less frequently considered alternatives are "intermediate," in the sense that they possess some of the traits 
of a black hole and some of the traits of a neutron star.  This possibility makes distinguishing the various possibilities 
even more difficult. 
\end{abstract}
\section*{Introduction}

Almost by definition, all supernovae leave an expanding remnant of their explosion, and these expanding 
remnants are the prime focus of this conference.  In addition at least some supernovae leave a 
condensed remnant which is of interest in its own right, and which may also strongly influence the larger expanding 
remnant.  We focus here on 
these condensed remnants.  Typically, three alternatives are considered: a neutron star, a 
black hole, or no remnant at all.  Type Ia Supernovae are thought to leave no remnant, but all other types are expected to leave 
either a black hole or a neutron star.  However, other types of collapsed remnants have been considered by 
many authors.  Among them are quark\cite{jgraber:qs1,jgraber:qs2}, strange\cite{jgraber:ss1,jgraber:ss2,jgraber:ss3,jgraber:ss4}
 and boson stars\cite{jgraber:bos1,jgraber:bos2}, soliton stars\cite{jgraber:sol1}, 
frozen stars\cite{jgraber:os39,jgraber:thorne,jgraber:rosen}, naked singularities\cite{jgraber:ns1,jgraber:ns2}, and 
wormholes\cite{jgraber:misner60,jgraber:bl63,jgraber:wh1}.  Other, less well-known, possibilities include  a type of object Mitra calls an 
eternally collapsing 
object(ECO)\cite{jgraber:mitra00},   and a 
somewhat similar object that I have discussed previously and
called a red hole\cite{jgraber:graber99}, but am now beginning to call a "big red bag," because this latter term is more descriptive. 
In this brief conference 
poster summary, we will not go into the theoretical motivations for considering these more exotic objects, nor 
into their technical definitions.  Instead, we will provide generic, phenomenological descriptions of these 
different types of object, and consider how they might be detected or rejected by observations of 
supernovae and especially of their condensed remnants.  In most cases, these same techniques are used 
to detect the difference between neutron stars and black holes, and to constrain certain properties of the 
condensed objects, such as neutron star equations of state.

The primary purpose of this paper is to urge that alternate models of condensed remnants not be overlooked, and to point out that 
standard models are already experiencing some difficulties explaining the observations.  Several significant 
observational constraints are already in hand and more are in the offing.   The second purpose is to suggest, 
and partly demonstrate, that both 
the theoretical models and the observational constraints can be expressed in terms of largely 
theory-free phenomenological parameters, for standard as well as more exotic objects. 

\section*{Tripartite Classification of Remnants}

Standard interpretation of standard theory, i.e., General Relativity (GR), leads us to expect only black holes or 
neutron stars.  We here consider in addition possibilities suggested by unusual forms of matter 
in standard GR, by nonstandard interpretations of GR, and by alternate theories of gravity.  In the spirit of the 
PPN approach to gravity,  we 
wish to consider in a categorical or parametric, theory-independent, way all possible condensed remnants, 
not just those that have already been considered in some detail.  Therefore we here consider not specific models, 
but classes of models, based on three generic characteristics that strongly affect the behavior, appearance, 
and detectability of the class of objects.

These three characteristics are the presence or absence of singularity(s), event horizon(s), and hard surface(s).  
Black holes and neutron stars differ on all three of these criteria.  Some of the other alternatives are "intermediate,"
 in that they possess some of the traits of a black hole and some of the traits of a neutron star.  This makes 
distinguishing the various possibilities even more difficult.

\section*{Classification of condensed objects}

We begin with the standard possibilities.  A black hole has an event horizon and a singularity, 
but no hard surface.  Just the opposite, a neutron star has a hard surface, but no event horizon 
and no singularity.  Among the less frequently considered alternatives, the strange, quark, boson 
and soliton stars all fall in the same category as the neutron stars; in fact they could be described
 as smaller, denser, and perhaps harder or stiffer neutron stars.  The naked singularity is in a class 
by itself, with no hard surface and no event horizon, but obviously a singularity.  Of course, the 
singularity can be pointlike or ringlike; spacelike, timelike or null; and of dimensions ranging from 
zero to three.  Many different varieties of naked singularity have been discussed.   Wormholes need to be subdivided 
to fit into our categories.  The classical wormhole\cite{jgraber:misner60,jgraber:bl63}, which has been described as "two black holes 
glued together at the event horizon," would have an event horizon, but no singularity and no hard surface;  the Einstein-Rosen 
bridge also falls in this category.  The more modern "traversable"   wormhole\cite{jgraber:wh1}, (which requires exotic 
matter for its construction) has no event horizon, no hard surface, and no 
singularity.  Also in this category are the red hole or the big red bag and the eternally 
collapsing object.   The frozen star\cite{jgraber:thorne} concept, which was an earlier understanding of the Schwarzschild 
solution that has now been largely replaced by the black hole paradigm, did seemingly have a hard surface, as well 
as what we now call an event horizon, and the singularity was hidden by both the event horizon and the 
hard surface. In this version, the frozen star had all three characteristics: a hard surface, an event horizon 
and a singularity.  In other interpretations (see e.g. Rosen\cite{jgraber:rosen}), the hard surface prevented the singularity 
from forming, and the frozen star  had only an event horizon and a hard surface 
without a singularity.

\section*{Observational effects of selected characteristics }

A hard surface can absorb or reflect infalling energy in the form of matter or 
radiation.  If it does absorb the infalling energy, it heats up and reradiates at least 
some of the energy, perhaps at a different wavelength and perhaps slowly over 
time, but all the infalling energy is at least potentially observable and, in principle, 
recoverable.  Thus accretion energetics and cooling curves can the presence or absence of a hard surface.

An event horizon is a "one-way membrane" that absorbs and hides all 
infalling energy.  The energy is lost from the view of the external observer 
and can not be seen or recovered (except quantum mechanically), although 
its mass can be detected gravitationally.  Hence energy balance calculations and 
observations can be a critical indicator of an event horizon.

A point singularity can be treated as a hard surface of radius zero that 
immediately reflects or reradiates all infalling energy.  Or it can be treated 
as an event horizon of radius zero that absorbs without (nongravitational) 
trace all infalling energy in whatever form.  Or it can be treated as a source of 
total unpredictability, leading to totally random and unpredictable results, possibly 
even including nonconsevation of normally conserved quantities.  Or it can be treated 
as an indication that the theory has broken down, and must be modified.  Higher dimensional 
and more complex singularities can be treated  analogously.

If the singularity is not hidden, it might be indicated by higher temperatures, faster re-radiation, and a smaller apparent size.

\section*{distinguishing collapsed remnants }

It has been surprisingly difficult  to detect the difference between black holes and neutron stars.  
So far the only ironclad technique has been the detection of highly regular pulsar radiation, which 
is conclusively diagnostic for the presence of a neutron star.  Other observational surprises include 
the inability to detect any compact remnant in many non-type Ia supernovae remnants, and the 
difficulty of conclusively detecting pulsars whose beam is not directed at us.  Detecting intermediate 
objects whose properties are less well understood can only be substantially more difficult.  Nevertheless, 
there have been observations that are hard to interpret with standard neutron star and black hole models,
which have led to the suggestion that perhaps one of these less familiar candidates is being observed.

Possible means of observing or constraining the condensed remnant include: direct and indirect observation 
of size and shape, collapse energetics, early and late cooling curves, and chemical composition of SNR ejecta.

\end{document}